\documentclass[twocolumn, pra,showpacs,superscriptaddress]{revtex4}
\usepackage{graphicx}
\usepackage{subfigure}
\usepackage{dcolumn}
\usepackage{bm}
\usepackage{amsmath}

\begin{document}

\title{Ground state properties of hard-core anyons in a harmonic potential}
\author{Yajiang Hao}
\email{haoyj@ustb.edu.cn}
\affiliation{Department of Physics,
University of Science and Technology Beijing, Beijing 100083, China}
\date{\today}

\begin{abstract}
Using anyon-fermion mapping method, we investigate the ground state properties of hard-core anyons confined in a one-dimensional harmonic trap. The concise analytical formula of the reduced one-body density matrix are obtained. Basing on the formula, we evaluated the momentum distribution, the natural orbitals and their occupation distributions for different statistical parameters. The occupation and occupation fraction of the lowest natural orbital versus anyon number are also displayed. It is shown that the ground state properties of anyons interplay between Bosons and Fermions continuously. We can expect that the hard-core anyons of larger statistical parameter exhibit the similar properties to the hard-core Bosons although anyon system satisfy specific fractional statistics.
\end{abstract}
\maketitle

\section{introduction}

Quantum statistical property is one of the most important properties of the particles in quantum system. According to the exchange symmetry satisfied by the wavefunction of identical particles, we can distinguish between fermions and bosons.  The former satisfy antisymmetry and the later symmetry under exchange. It is the case that the identical particles exchange in three dimensions (and higher dimensions) systems.  In one- and two-dimensional systems this respect is qualitatively different from the high-dimensional systems \cite{anyons}. Physicists proposed that there exists a continuum of intermediate cases between Bosons and Fermions, i.e., the anyons  satisfying the fractional statistics \cite{Wilczek}. The fractional statistics was used successfully to explain the  fractional quantum Hall effect \cite{SuperConductor} and nowadays play important roles in condensed matter physics\cite{Wilczek,Laughlin,Halperin,Camino,YSWu,ZNCHa}. One interest feature of fractional statistics system is the application to quantum information science  for the topological protection of quantum coherence of anyons \cite{RMP2008,Kitaev}. It is the feature that catalyzed the search for the anyons in physical systems such as quantum Hall heterostructure \cite{Mong} and low dimensional quantum gas.

For the high controllability quantum gas is a popular platform to simulate the traditional condensed matter system. Several ideas have suggested the creation of anyons with rotated Bose-Einstein condensates (BECs) \cite{PZoller} and cold atoms  in optical lattices \cite{DuanLM,OL}. Besides the realization and manipulation of anyons in two-dimensional optical lattice, theorists proposed that the one-dimensional (1D ) anyon gas of tunable interaction can be created in 1D optical lattice by controlling the transition rate with the Raman-assisted hopping technique \cite{NatureComm,LSantos}. Recently, Str\"{a}ter etc. proposed a new mechanism to create anyons in optical lattice by controlling the lattice-shaking-induced tunneling \cite{Strater}. The 1D quantum gas is one of the focuses in cold atom physics both theoretically \cite{1D} and experimentally \cite{Ketterler,Paredes,Toshiya} since the realization of the Bose-Einstein condensates. Particularly, with Feshbach resonance technique the strong correlated Tonks-Girardeau (TG) gas \cite{Paredes,Toshiya,Jacqmin} and super TG gas \cite{STG} are achievable. The atomic interaction can be tuned in the full interacting regime with the Feshbach resonance technique and confinement-induced resonance technique. Most interestingly, the mechanism proposed by Greschner etc. can control the effective interaction between anyons without Feshbach resonances \cite{LSantos}.

Although the experimental realization of 1D anyon gas in cold atom is still of absence, its theoretical research has been in progress
for a long time \cite{Haldane,WangZD,Kundu99,Girardeau06}. Particularly the $\delta$-anyon gas attracted many research interests in the exact solution \cite{Kundu99,Girardeau06,Batchelor}, the low-energy properties \cite{XWGuanLowEnergy}, correlation function \cite{Patu07,Patu08,Calabrese}, entanglement properties \cite{Cabra,HLGuo}, momentum distribution and the reduced one-body density matrix (ROBDM) \cite{Cabra,anyonTG,HaoPRA78,HaoPRA79}, the relaxation dynamics \cite{MRigol}, quantum walks \cite{YZhang} and the fermionization \cite{Campo,HaoPRA2012}. It turns out that the properties dependent on the modulus of wavefunction are not related with the statistical properties, while those properties dependent on the wave function exhibit distinct behaviors with the change of the statical parameters. For example, the density distribution and pair correlation function displays the same behavior as those of Bose gas \cite{HaoPRA78,HaoPRA79}. The properties resulted from the fractional statistics are exhibited for ROBDM and the momentum distributions \cite{Cabra,anyonTG,HaoPRA78,HaoPRA79,Campo,HaoPRA2012}. The asymmetric momentum distribution of anyons is greatly different from the symmetric distribution of fermions and bosons.

For homogeneous anyon gas the exact solution and therefore the interested properties can be obtained with Bethe ansatz method. For the anyon gas in external potential we have to turn to numerical method except that in the strong interaction limit. The strongly interacting anyons can be mapped into spin-polarized fermi gas with anyon-fermion mapping method and the  generalized Jordan-Wigner transformation for the anyons in harmonic potential and in optical lattice, respectively. Hao etc. have obtained the exact ground state of strongly interacting anyon gas in optical lattice \cite{HaoPRA79}. The ground state of strongly interacting anyon gas in a harmonic trap was also investigated by Girardeau \cite{Girardeau06} where the system was discussed formally, while the ROBDM and thus the density profiles and the momentum distribution were not investigated. In the present paper we will give the exact analytical formula of ROBDM and other related quantities. The formula can be used to investigate the system of large particle number.

The paper is organized as follows. In Sec. II, we give a brief review of 1D anyonic model and introduce the analytical solution. In Sec. III, we present the ROBDM, momentum distributions, occupation distribution and the lowest natural orbital for different statistical parameters. A brief summary is given in Sec. IV.

\narrowtext
\section{model and method}
We consider $N$ anyons with the hard-core interaction trapped in a harmonic potential
\begin{equation}
V_{ext}=x^{2}/2.
\end{equation}
For simplification the natural unit is used here and in the following formula. For the hard-core interacting anyon gas, its wavefunction can be obtained by the wavefunction of the polarized fermi gas with the anyon-fermion mapping method
\begin{equation}
\Phi_{A}(x_{1},\cdots,x_{N})=\mathcal{A}(x_{1},\cdots,x_{N})\Phi_{F}\left(x_{1},x_{2},\cdots,x_{N}\right).
\end{equation}
Here the anyonic mapping function is formulated as
\begin{equation}
\mathcal{A}(x_{1},\cdots,x_{N})=\prod_{1\leq j<k\leq N}\exp[-\frac{i\chi \pi}{2}\epsilon(x_j-x_k)]
\end{equation}
with $0 \le \chi \le 1$ being the statistical parameter. 1 corresponds to the hard-core Bosons and 0 corresponds to noninteracting fermions. The wavefunction of $N$ polarized fermions $\Phi_{F}\left(x_{1},x_{2},\cdots,x_{N}\right)$ can be constructed by the one-particle wavefunction as
\begin{equation}
\Phi_{F}\left(x_{1},x_{2},\cdots,x_{N}\right)=\left(1/\sqrt{N!}\right)\det_{j,k=1}^{N}\phi_{j}\left(x_{k}\right).
\end{equation}
Here $\phi_{j}(x)$ is the $j$th eigen wavefunction of one particle in harmonic trap $\phi_{j}(x)=(\sqrt{\pi}2^{j}j!)^{-\frac12}e^{-x^{2}/2}H_{j}(x)$. Using the following properties of the Vandermonde determinant formula
\begin{equation}
\det[p_{j-1}(x_{k})]_{j,k=1,\cdots,N}=\prod_{1\leq j<k\leq N}(x_{j}-x_{k})
\end{equation}
for $\{p_{j}(x)\}$=$\{2^{-j}H_{j}(x)\}$, the exact many body wavefunction of $N$ polarized fermions can be formulated as
\begin{eqnarray}
&&\Phi_{F}\left(x_{1},x_{2},\cdots,x_{N}\right)    \\ \nonumber
&&=(C_N^H)^{-1}\prod_{j=1}^{N}exp(-x_{j}^{2}/2)\prod_{1\leq j<k\leq N}(x_{j}-x_{k})
\end{eqnarray}
with $(C_N^H)^{-2}=\pi^{-N/2}N!^{-1}\prod_{j=0}^{N-1}2^{j}j!^{-1}$. Therefore the wavefunction of hard-core anyons should be
 \begin{eqnarray}
\Phi_{A}\left(x_{1},x_{2},\cdots,x_{N}\right)=\mathcal{A}(x_{1},x_{2},\cdots,x_{N})   \\	\nonumber
\times \frac{1}{C_{N}^{H}}\prod_{j=1}^{N}exp(-x_{j}^{2}/2)\prod_{1\leq j<k\leq N}(x_{j}-x_{k}).
\end{eqnarray}

With the above many body wavefunction of anyons, the reduced one body density matrix (ROBDM) of anyon gas can be calculated by
\begin{eqnarray}
\rho(x,y)&=&N\int_{-\infty}^{\infty}dx_{1}\cdots\int_{-\infty}^{\infty}dx_{N-1}		\\	\nonumber
&&\times \Psi_{A}^{*}(x_{1},\cdots,x_{N-1},x)\Psi_{A}(x_{1},\cdots,x_{N-1},y).
\end{eqnarray}
Substituting the wavefunction  Eq. (7) into the integral formula, we find that the ROBDM takes the form of determinant of Hankel type
\begin{eqnarray}
\rho(x,y)&=&\frac{N}{(C_{N}^{H})^{2}}exp(-x^{2}/2-y^{2}/2)\prod_{j=1}^{N-1}\int_{-\infty}^{\infty}dx_{j}	\\	\nonumber
&& \times exp(-x_{j}^{2})e^{i\chi\pi\epsilon(x_{j}-x)/2}(x_{j}-x)e^{-i\chi\pi\epsilon(x_{j}-y)/2}			\\	\nonumber
&& \times (x_{j}-y)(\det[x_{k}^{j-1}]_{j,k=1,\cdots,N-1})^{2}.
\end{eqnarray}
With the properties of determinant we get the concise expression of ROBDM
\begin{eqnarray}
\rho\left(x,y\right)&=&\frac{2^{N-1}}{\sqrt{\pi}\Gamma\left(N\right)}exp(-x^{2}/2-y^{2}/2)	\\	\nonumber
&& \times \det[\frac{2^{(j+k)/2}}{2\sqrt{\pi}\sqrt{\Gamma\left(j\right)\Gamma\left(k\right)}}b_{j,k}\left(x,y\right)]_{j,k=1,\cdots,N-1}
\end{eqnarray}
with
\begin{eqnarray}
&&b_{j,k}\left(x,y\right)=		\\	\nonumber
&&\int_{-\infty}^{\infty}dtexp(-t^{2})e^{i\chi\pi(\epsilon(t-x)-\epsilon(t-y))/2}(t-x)(t-y)t^{j+k-2}.
\end{eqnarray}
For convenience $b_{j,k}(x,y)$  can be reformulated  as the integral expression in different integral range with special functions, {\it i.e.},
\begin{eqnarray}
&&b_{j,k}(x,y)=		\\	\nonumber
&&f_{j,k}(x,y)+(e^{i\chi\pi\epsilon(y-x)}-1)\epsilon(y-x) [xy	\\	\nonumber
&&\times \mu_{j+k-2}(x,y) -(x+y)\mu_{j+k-1}(x,y)+\mu_{j+k}(x,y)].
\end{eqnarray}
Here the function $f_{j,k}(x,y)$ depends on Gamma function and $\mu_{m}(x,y)$ depends on confluent hypergeometric function \cite{Forrester}.

The diagonal part of ROBDM is the density distribution in real space. According to Eq. (9), the density profiles of hard-core anyons are independent on the statistical properties, which are exactly same as those of hard-core bosons and spin-polarized fermions. The momentum distribution is defined as its Fourier transformation
\begin{eqnarray}
n(k)=\frac{1}{2\pi}\int_{-\infty}^{\infty}dx\int_{-\infty}^{\infty}dy\rho(x,y) e^{-ik(x-y)}.
\end{eqnarray}
In the present paper we shall also investigate the natural orbital of the hard-core anyons, which can be understood as the effective single-particle states. The natural orbitals $\varphi _{\eta}(x)$ are defined as the eigenfunctions of the one-particle density matrix
\begin{eqnarray}
\int_{-\infty}^{\infty}dy\rho(x,y)\varphi  _{\eta}(y)=\lambda _{\eta}\varphi  _{\eta}(x),
\eta =1,2,...,
\end{eqnarray}
where $\lambda _{\eta}$ is the occupation of the $\eta$th natural orbital $\varphi _{\eta}(x)$. Obviously, the ROBDM, the momentum distribution, the natural orbital and its occupation number are dependent on the statistical parameter $\chi$.

\section{Ground state properties of 1D hard-core anyon gases}
In this section,  using the analytical formula given in the previous section, we evaluate the ground state properties of 1D strongly interacting anyon gases of $N$ anyons in a harmonic trap for different statistical properties.

\begin{figure}
\includegraphics[width=3.0in]{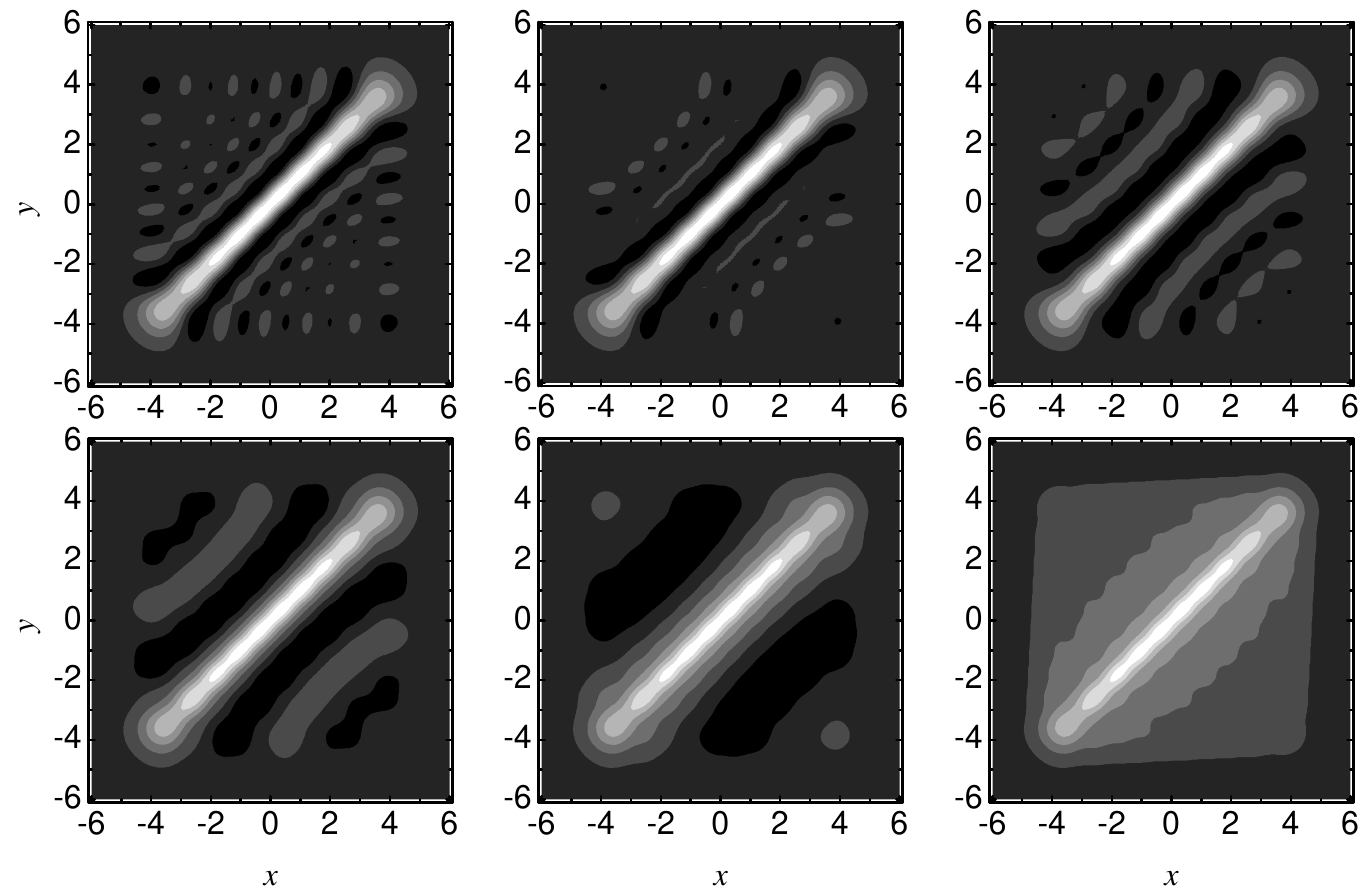}
\caption{The real part of ROBDM for anyon gas of $N$=10. The statistical parameter $\chi$= 0, 0.2, 0.4, 0.6, 0.8 and 1.0 (from left to right and from above to bottom).}
\end{figure}
\begin{figure}
\includegraphics[width=3.0in]{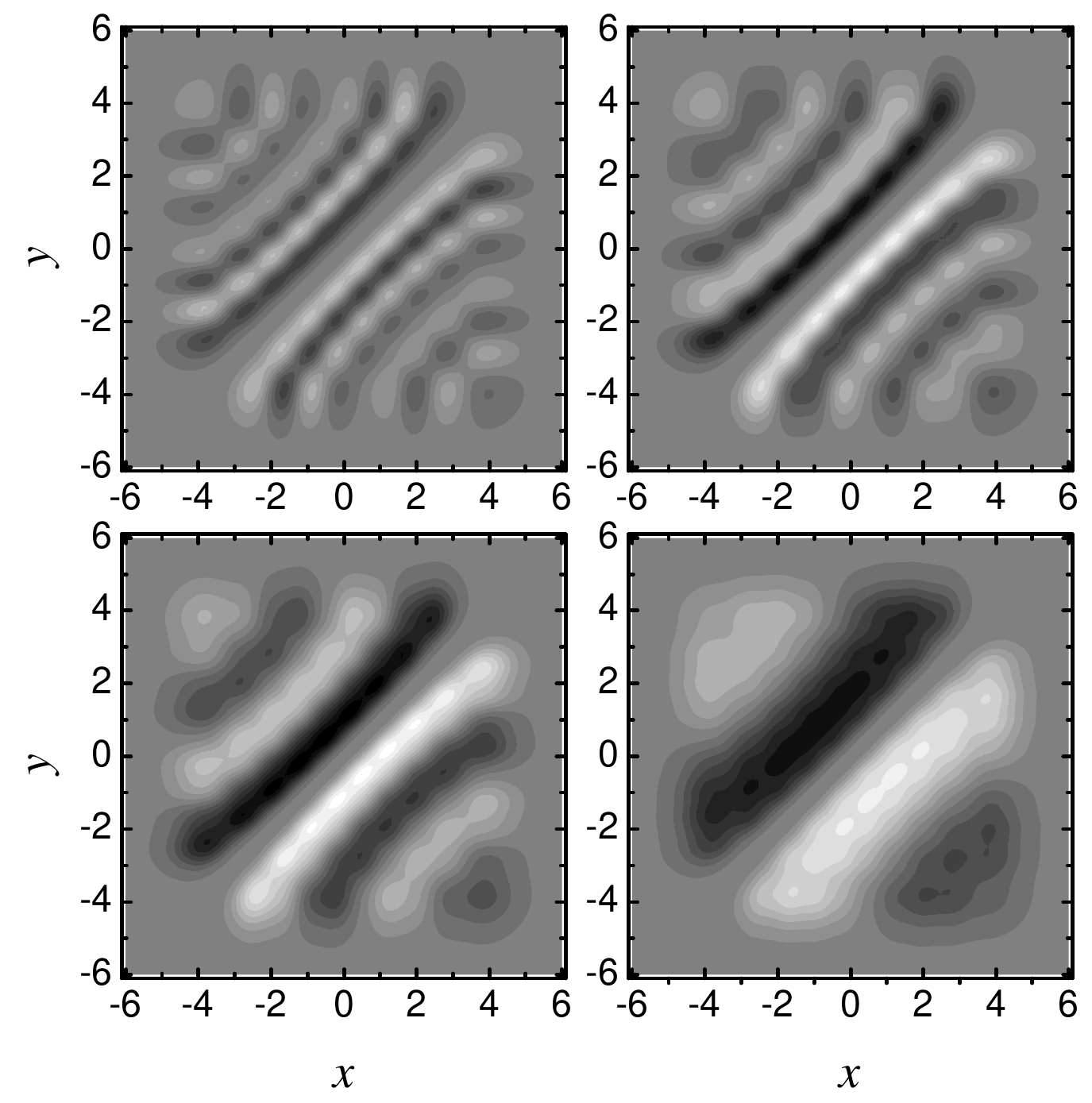}
\caption{The imaginary part of ROBDM for anyon gas of $N$=10. The statistical parameter $\chi$= 0.2, 0.4, 0.6 and 0.8 (from left to right and from above to bottom).}
\end{figure}
According to Eq. (10) the ROBDM is complex rather than real as that of TG gases. In Fig. 1 we display the gray scale plot of the real part of ROBDM Re[$\rho(x,y)$] of 1D anyon gas of $N=10$ for statistical parameters $\chi$= 0, 0.2, 0.4, 0.6, 0.8 and 1.0. The diagonal parts, i.e., the density distribution of anyon gases in the real space, exhibit the same behaviors irrelevant to the statistical parameters, which can also be verified by Eq. (9). For Bose statistics ($\chi$=1.0), the off-diagonal elements of ROBDM are not negligible, which embody the off-diagonal long range order \cite{Vaidya} and the properties of 1D Bose gas as a quasi-condensate. With the evolution of statistical properties from Bose statistics to Fermi statistics, the off-diagonal elements of ROBDM decrease in contrast to the diagonal part and finally become negligible small as $\chi=0.0$ (Fermi statistics). The imaginary parts of ROBDM are displayed in Fig. 2 for statistical parameters $\chi$= 0.2, 0.4, 0.6 and 0.8. For Bose statistics and Fermi statistics, the imaginary part of ROBDM is zero, while for anyon gases Im[$\rho(x,y)$] are antisymmetric about the diagonal line $x=y$. It is the nonvanishing imaginary part of ROBDM that result in the asymmetric momentum distribution of anyon gases.

\begin{figure}
\includegraphics[width=3.0in]{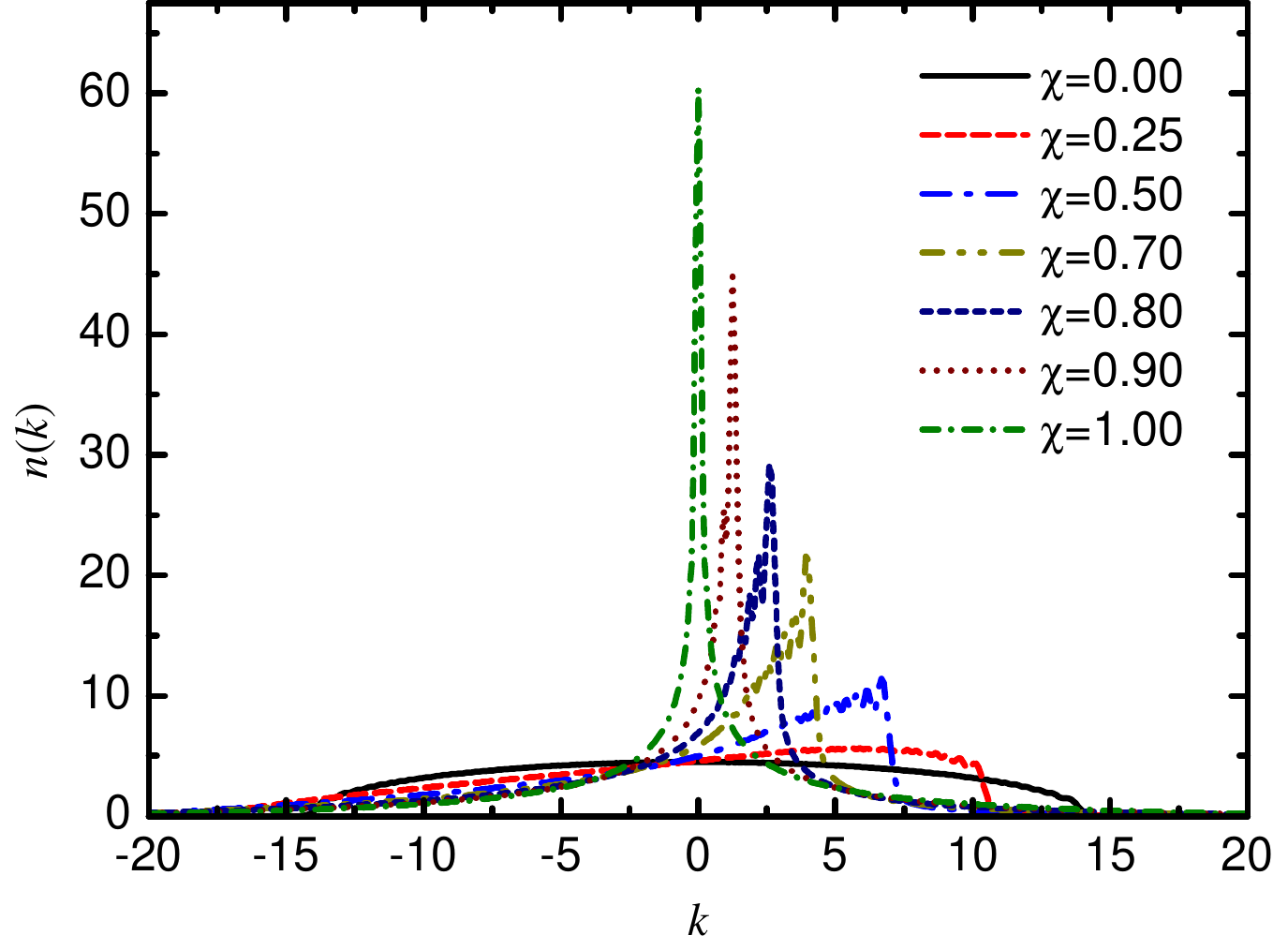}
\caption{Momentum distribution for anyon gas of $N$=100.}
\label{default}
\end{figure}

The momentum distributions of anyon gases are shown in Fig. 3 for different statistical parameters. For Bose statistics the momentum distribution exhibit $\delta$-function like profiles. Bose particles distribute in the regime of zero momentum with a great probability. With the increase of momentum the occupation probability decrease rapidly. Fermi particles distribute in a large momentum regime with almost equal probability. Compared with the Bose gas and Fermi gas, the remarkable feature of momentum distribution of anyons is the asymmetric profile. As the statistical parameters deviate from the Bose statistics, although momentum profiles still show $\delta$-function like sharp peak distribution in a small momentum regime, the most probable momentum will not be zero ($\chi=0.9$ for example). At the same time, the maximum occupation probability  decrease and the width of momentum profiles increase with the decrease of statistical parameter.

\begin{figure}
\includegraphics[width=3.5in]{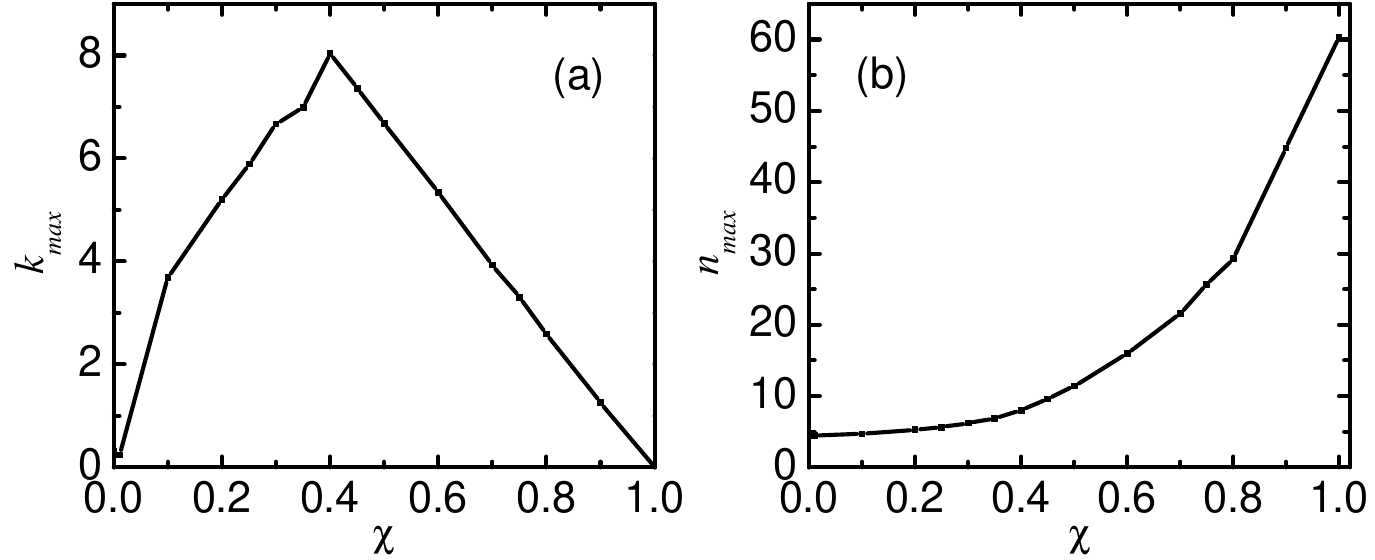}
\caption{The maximum occupation probability (b) and the most probable momentum (a) versus statistical parameter for different statistical parameters for anyon gas of $N$=100.}
\label{default}
\end{figure}
In Fig. 4 the maximum occupation probability $n_{max}$=Max[$n(k)$] and the most probable momentum $k_{max}$ are displayed for different statistical parameters. It is shown that the maximum deviation of $k_{max}$ from zero is arrived at for statistical parameter $\chi$=0.4. In the regime $\chi \ge 0.4$ the most probable momentum depends on the statistical parameter linearly. As $\chi < 0.4$ the linear statistical parameter dependence of $k_{max}$ is broken because the obvious sharp peak of momentum distribution vanishes. The height of momentum profiles is maximum for hard-core bosons. It should be noticed that the total momentum (and therefore the mean momentum) of anyons is zero rather than a finite value although its distribution is not symmetric about zero momentum, which is verified by evaluating $k_{total}=\int_{-\infty}^{\infty}kn(k)dk$.

The eigenvalues of ROBDM $\lambda_i$ are occupation number of natural orbitals. If the largest occupation number is macroscopic, the system will exhibit quasi-BEC and the corresponding natural orbital play a role of an order parameter \cite{Girardeau06}. The occupation distribution of natural orbital $\lambda_i$ versus orbital number $i$ for 100 anyons are displayed in Fig. 5. It is shown that anyons with larger statistical parameter exhibit the similar properties to that of Bosons. The anyons occupy macroscopically the lowest natural orbital $\varphi_0(x)$  and the occupation number decrease significantly for higher-order orbitals. With the decrease of statistical parameter more and more higher-order natural orbitals are occupied and finally the occupation distribution behave similar to that of spin-polarized fermions, in which fermions occupy the $N$ lowest natural orbitals and occupation number is 1. Therefore we can expect the lowest natural orbital dominate the properties of anyons with larger statistical parameter, and this will be problematic for anyons with small statistical parameter.

\begin{figure}
\includegraphics[width=3.0in]{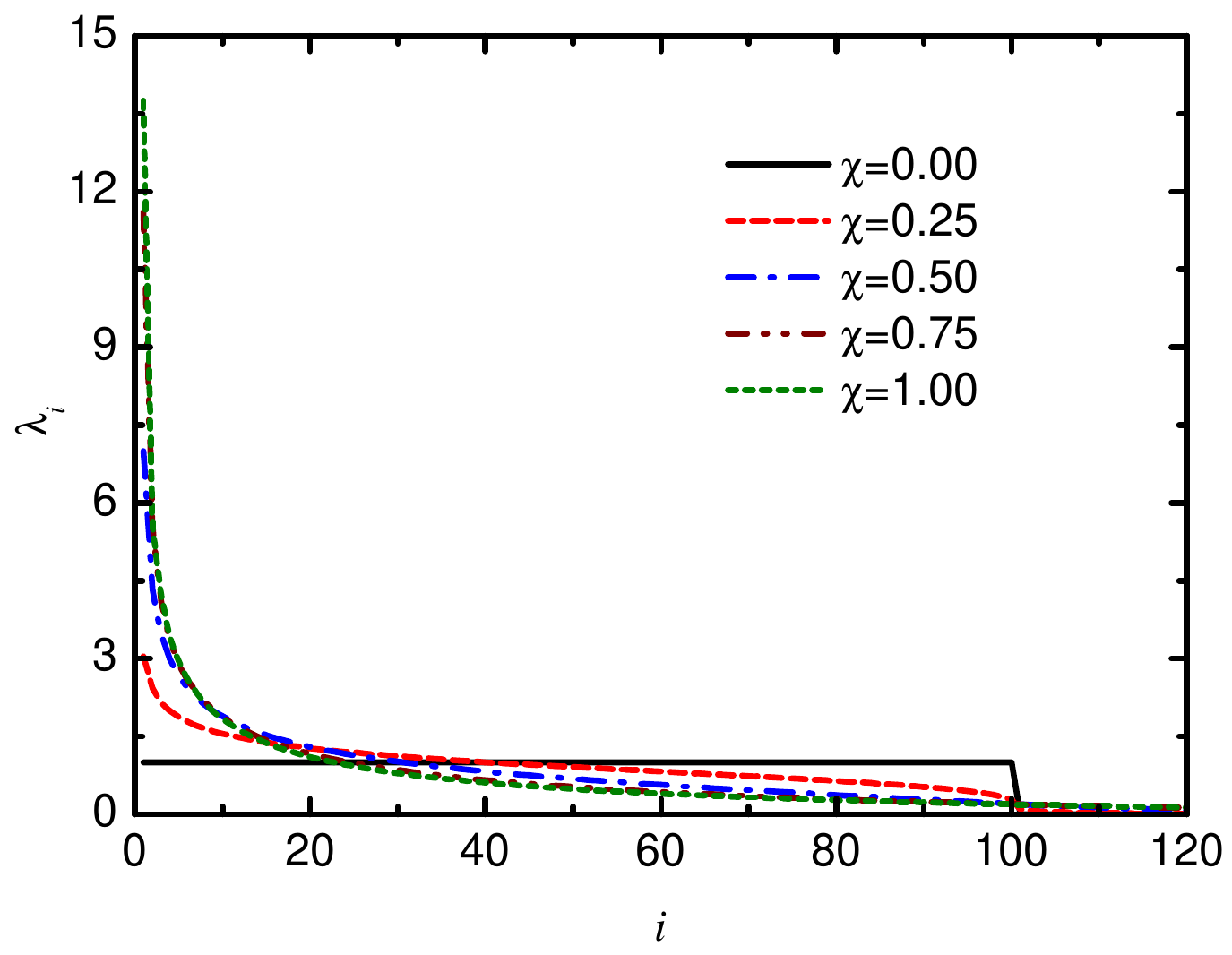}
\caption{The occupation of the natural orbitals for different statistical parameters for anyon gas of $N$=100.}
\label{default}
\end{figure}

The occupation number of the lowest natural orbital $\lambda_0$ versus anyon number are displayed in Fig. 6a. As the increase of anyon number more anyons occupy in the lowest natural orbital except the case of fermi statistics. The occupation can be fitted to the power-law $\lambda_0\approx AN^{\alpha}$ with $(A, \alpha)$=(1.169, 0.538), (1.154, 0.502), (1.113, 0.400), (1.061, 0.230) and (1.000, 0) for statistical parameter $\chi$=1.0, 0.75, 0.5, 0.25 and 0.0, respectively. The fitting power law are plotted in short dot lines in Fig. 6a. We also display the occupation fraction of the lowest natural orbital $\lambda_0/N$ versus the total anyon number. For anyons the occupation fraction decreases with increasing anyon number $N$, but the slope become lower for large anyon number.

\begin{figure}
\includegraphics[width=3.5in]{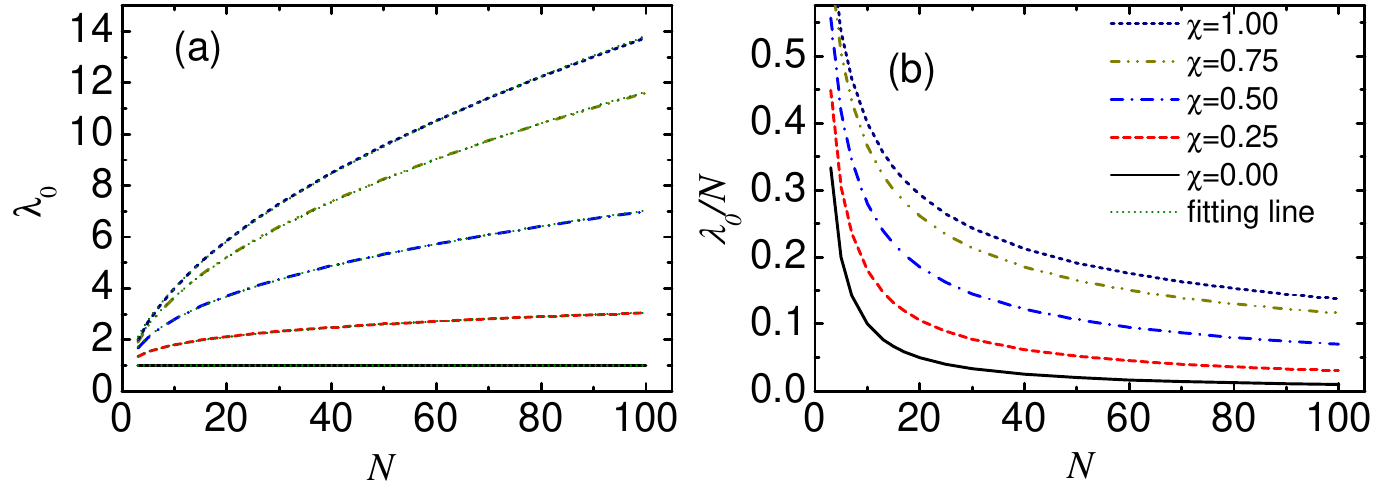}
\caption{The occupation number $\lambda _0$ (a) and occupation fraction $\lambda _0/N$ (b) in the lowest natural orbital versus particle number $N$ for different statistical parameters. In (a) the fitting lines $\lambda_0\approx AN^{\alpha}$ are plotted in short dot lines.}
\label{default}
\end{figure}

\begin{figure}
\includegraphics[width=2.5in]{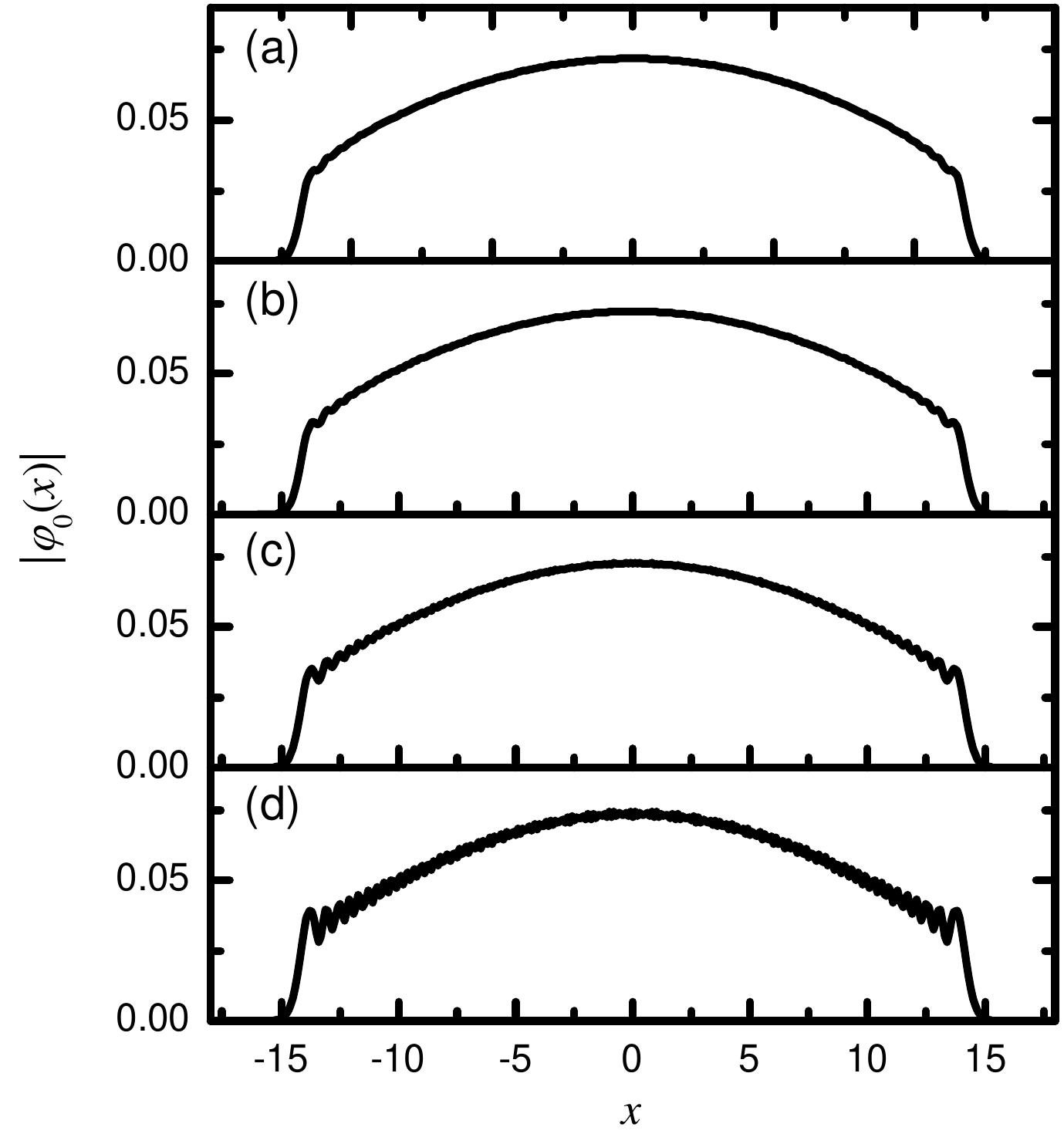}
\caption{The module of the lowest natural orbital for anyon gas of $N$=100. (a) $\chi=1.0$; (b) $\chi=0.75$; (c) $\chi=0.5$; (d) $\chi=0.25$.}
\label{default}
\end{figure}

The modulus of the lowest natural orbital $|\varphi_0(x)|$ are shown in Fig. 7 for different statistical parameter. The natural orbital of anyons exhibits the similar profiles for different statistical parameter. With the decreasing statistical parameter more oscillation are shown and as the statistical properties are close to Fermi statistics the oscillation become drastic.

\section{Summary}
In conclusion, we have obtained the analytical formula of ROBDM for the ground state of 1D hard-core anyons in a harmonic potential with the anyon-Fermion mapping method. Basing on the formula, we evaluated the momentum distribution, the natural orbitals and occupation distribution for different statistical properties. The ground state properties evolve continuously from Bose-like to Fermi-like with the decrease of statistical parameter.

For larger statistical parameter, the momentum distribution display a $\delta$-function like sharp peak profiles at a finite momentum, which is similar to that of hard-core Bosons. With the decreasing statistical parameter, the sharp peak of momentum profiles become lower and broader, and the position of peak deviates from the zero momentum firstly and then approach to zero. The maximum deviation of the peak position is arrived at for $\chi=0.4$. The occupation distribution of natural orbitals also display the behaviours interpolating between the Bose-like properties and fermi-like properties for different statistical parameters. In the Bose case, the lowest natural orbital is occupied macroscopically and higher natural orbitals are occupied with a minimal probability. In the Fermi case, the occupation number is one for each natural orbital. We can expect that the anyon gas of larger statistical parameter display the similar properties to bose gases. The occupation of the lowest natural orbital behave a power-law relation dependent on statistical parameter with anyon number.

\begin{acknowledgments}
This work was supported by NSF of China under Grants No. 11004007 and ``the Fundamental Research Funds for the Central Universities."
\end{acknowledgments}

\end{document}